\documentstyle[aaspp4,11pt,epsf]{article}

\begin{document}

\clubpenalty = 10000
\widowpenalty = 10000
\title{Faint Field Galaxies Around Bright Stars - A New Strategy
for Imaging at the Diffraction Limit}
\author{J. E. Larkin \& T. M. Glassman}
\affil{Dept. of Physics and Astronomy, University of California, Los Angeles}
\authoraddr{Dept. of Physics and Astronomy, 8967 Math Sciences,
Los Angeles, Ca. 90095-1562}
\authoremail{larkin@astro.ucla.edu}


\begin{abstract}

This paper presents a new strategy for observing faint galaxies with
high order natural guide star systems. We have imaged 5 high galactic
latitude fields within the isoplanatic patch of bright stars (8.5 $<$
R $<$ 10.3 mag).  The fields provide a rich set of faint field
galaxies that are observable with a natural guide star adaptive optics
system on a large telescope.  Due to the small fields of many AO
science cameras, these preliminary images are necessary to identify
candidate galaxies.  We present the photometry and positions for 78
objects (at least 40 galaxies) near five bright stars, appropriate for
diffraction limited studies with the Keck and other AO systems on
large ground-based telescopes.  The K band seeing conditions in each
field were excellent (0\farcs4 - 0\farcs7) allowing us to identify
stars and estimate galaxy sizes.  We also simulate AO images of field
galaxies to determine the feasibility of infrared morphological
studies at the diffraction limit.  With new high order AO systems
coming on line with 8-10 meter class telescopes, we believe these
observations are invaluable in beginning to study faint galaxy populations
at the diffraction limit.

\end{abstract}


\section{Introduction}

It has only been very recently, with the help of the Keck Telescopes
and the Hubble Space Telescope (HST) that galaxies have been
identified which are thought to be producing their first generation of
stars.  With the high resolution of the HST many of these young
galaxies appear to be more numerous and smaller than nearby galaxies
(Phillips et al. \markcite{phi} 1997), and often have a distorted
morphology (Driver et al. \markcite{dri} 1998).  Taken together,
these attributes suggest that galaxies have gone through a period of
significant evolution since their formation.

Optical imaging, however, is biased by the fact that at high redshifts
the observed light was emitted in the UV where star forming regions
dominate the emission. This can result in a more distorted appearance
and give a biased estimate of the morphology.  It is probable then,
that at least some of the close groupings of optical knots seen in
deep HST images may actually be multiple star forming regions within a
single galaxy.  Infrared cameras can directly image the
optical emission from high redshift galaxies and provide a more
accurate determination of the galaxy's morphology.  With the NICMOS
camera on HST, images of the Hubble Deep Field (Thompson et al.
\markcite{tho} 1999) have shown that for at least some objects, their
infrared morphology is in fact smoother and less complex than their
optical morphology.  High redshift objects are also very small,
usually less that one arcsecond in extent, so direct ground based
images often yield little morphological information.  Even with
NICMOS resolutions are limited to about 0$\farcs$2 barely resolving
many galaxies.  What is needed is diffraction limited observations from
larger (8-10 m class) ground based telescopes in the infrared.

Adaptive Optics (AO) Systems coupled to new and anticipated infrared
instruments will be able to probe the infrared morphologies of distant
galaxies in much more detail than previous studies. An intrinsic
problem with the earliest form of most high order AO systems, however,
is their reliance on bright natural guide stars; often brighter than
about 12th magnitude at R. This limitation makes most extragalactic
targets unobservable because of their intrinsic faintness and the
relative rarity of sufficiently bright nearby guide stars.  Recent
observations with relatively low Strehl ratios of a few quasars and
radio galaxies have been possible with curvature type AO systems due
to the ability of these systems to operate at slower speeds and larger
effective aperture (Stockton et al.  \markcite{scm} 1999, Hutchings et
al. \markcite{hcm} 1999, and Aretxaga et al. \markcite{alm}
1998). Laser guide stars will partially remedy this problem in a
couple of years, but laser systems usually produce lower Strehl ratios
than natural guide star systems, so sensitivities and resolution will
suffer.

We have developed an interesting new strategy for using natural guide star
systems to observe extremely faint galaxies. These observations rely
on the high density of galaxies on the sky. Deep infrared surveys
(e.g. Djorgovski et al. \markcite{djo} 1995) have shown that there
are about 2x10$^5$ galaxies per square degree brighter than K=24 mag.
To a limiting magnitude of K=20 this number is down by about a
factor of ten to 2x10$^4$ per square degree, but this still implies
that within 20 arcseconds of ANY guide star there are on average 2
galaxies brighter than K=20 mag and 20 galaxies brighter than 24th
magnitude. Recent redshift surveys (e.g. Cohen et al. \markcite{coh}
1996) have shown that the average redshift of field galaxies brighter
than K=20 mag is greater than 0.5 and this should rise at fainter
magnitudes. So our strategy is to perform deep infrared imaging around
bright ($<$12 mag) A-type stars to identify faint galaxies, then use
the much smaller field of view of the AO infrared cameras to image
selected galaxies with high Strehl ratio at or close to the
diffraction limit.

We present here our first infrared images near five bright stars that
are at relatively high galactic latitude, have a relatively blue color
(A spectral type), and which pass close to the zenith of the Keck
Observatory and other Northen Hemisphere Observatories.  We calculate
infrared colors for two of the fields, and crude morphologies when
possible to allow for better early selection of potentially
interesting objects.

\begin{deluxetable}{lcccccccc}
\small
\tablewidth{0pt}
\tablecaption{\bf List of Stars Observed}

\tablehead{\colhead{} &\colhead{} &\colhead{} &\colhead{} &
\colhead{Spectral} & \colhead{} & \colhead{Band \&} 
& \colhead{Seeing} & \colhead{} \nl
\colhead{Object} & \colhead{RA(2000)} 
& \colhead{Dec(2000)} & \colhead{R (mag)} 
&\colhead{Type} & \colhead{gb} & \colhead{Exp.Time} 
& \colhead{FWHM} & \colhead{Date} \nl
\colhead{(1)} & \colhead{(2)} & \colhead{(3)} 
& \colhead{(4)} & \colhead{(5)}& \colhead{(6)} 
& \colhead{(7)} & \colhead{(8)} & \colhead{(9)}}

\startdata
ppm 91088 & 01 48 02.12 & +21 48 28.48 & 10.2 & A3 & -39 11 & K - 3060 s &
0\farcs4 & 10Oct98 \nl
ppm 91714 & 02 28 48.09 & +23 52 13.67 & 8.5 & A2 & -33 51 & K - 1080 s &
0\farcs5 & 06Sep98 \nl
& & & & & & K - 540 s & 0\farcs4 & 08Oct98 \nl
& & & & & & H - 540 s & 0\farcs5 & 08Oct98 \nl
& & & & & & J - 1080 s & 0\farcs8 & 08Oct98 \nl
ppm 50296 & 08 02 10.51 & +42 28 29.3 & 9.9 & A5 & +30 29 & K - 1620 s &
0\farcs4 & 10Oct98 \nl
ppm 98537 & 08 16 30.93 & +22 35 03.17 & 10.2 & A0 & +28 15 & K - 1620 s &
0\farcs7 & 08Oct98 \nl
ppm 106365 & 17 47 59.48 & +22 40 14.37 & 10.3 & A2 & +23 35 & K - 1620 s &
0\farcs5 & 06Sep98 \nl
& & & & & & H - 1080 s & 0\farcs5 & 06Sep98 \nl
& & & & & & J - 1080 s & 0\farcs6 & 06Sep98 \nl
\enddata
\label{t:irlog}
\end{deluxetable}

\section{Observations}

We've used the Keck Near Infrared Camera (NIRC) to image around a
sample of five early-type stars with visual magnitudes between 8.5 and
10.3. A-type stars are preferred because they are relatively blue and
thus reduce the amount of scattered and diffracted light in the
infrared images as compared to other stars of comparable optical
magnitude. O and B type stars are of course bluer, but are relatively
rare at high galactic latitudes.  The stars were also selected to have
low proper motions ($<$0\farcs01 per year), relatively high galactic
latitude (lb$>$20$^o$ or lb$<$-20$^o$), and a declination within 5
degrees of Keck's latitude (but not those passing through the zone of
avoidance near the zenith); at certain RA, the galactic
latitude constraint forced us further north.

Table 1 gives the list of stars observed along with their coordinates,
R band magnitudes, spectral types and galactic latitudes.  Also given
are the infrared exposure times for each band used (J, H or K), the
spatial resolution of the final summed infrared images and the date of
the observations.  All observations were performed with the Near
Infrared Camera (NIRC, Matthews \& Soifer \markcite{mat} 1994) on the
Keck I Telescope on the nights of 06 September, 1998 and 08-10
October, 1998. Conditions on each night were clear and photometric.
Typical seeing was between 0.4 and 0.7 arcseconds, but was as good as
0\farcs2 and as bad as 1\farcs0 during certain short periods.

In each band, many individual frames were taken. For the K band, each
frame consisted of 20 coadded exposures of 3 seconds each, except for
the observation of PPM91714 on 08 October, 1998 which had 60 coadded
exposures of 1 second each.  For H, each frame consisted of 60 coadded
exposures of 1 second each, and for J, they were 12 exposures of 10
seconds each.  Image sequences consisted of a 3 by 3 pattern with a
step size of 5'' along each axis.  This yielded 9 minutes of exposures
for each sequence, except for the J band sequences which were
18 minutes for each 3 by 3 grid.  For deeper observations, additional
3 by 3 sequences were taken with a small (typically 3'') offset
between sequences.  Because of the magnitudes of the stars themselves,
they always saturated and were positioned at the corner of the array
to reduce the effect of electronic bleeding and diffraction spikes
within the images.  This has the drawback of reducing the amount of
the isoplanatic patch that was covered.

The data were reduced with custom IDL routines that medianed images in
groups of 9 without aligning the frames in order to make sky and flat
fields.  Bright objects were masked from the images before producing
the skies and flats.  Each sky and flat was only used on the central 3
images within the group of 9.  This produced very good skies and flats
that accurately match the varying sky levels throughout the
observation without reducing the observing efficiency by taking
separate sky frames.  The sky subtracted and flat fielded frames are
then aligned to the nearest integer pixel and combined using a clipped mean at
each pixel.  The final images have very uniform backgrounds and noise
consistent with the square root of the number of frames used in each
mosaic.

\clearpage
\section{Results}

Figures 1 through 5 show the images of the stellar fields.  They
have been stretched very hard to show the faint galaxies in the
central, cleanest, parts of the images; this makes the noise at the
edges appear artificially extreme.  In each case, the noise is
consistent with the number of frames contributing at that pixel, except
where diffraction spikes or bleeding leave residual images.

\begin{figure}[htb]
\plotfiddle{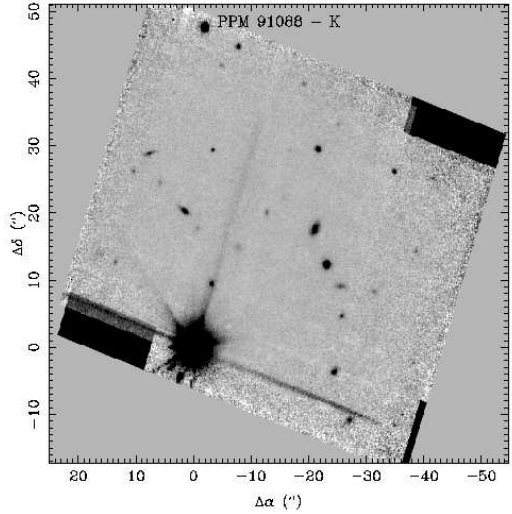}{2.4in}{0}{40}{40}{-120}{-70}
\caption{K band image of the field around ppm 91088.}
\end{figure}

\begin{figure}[hbt]
\plotfiddle{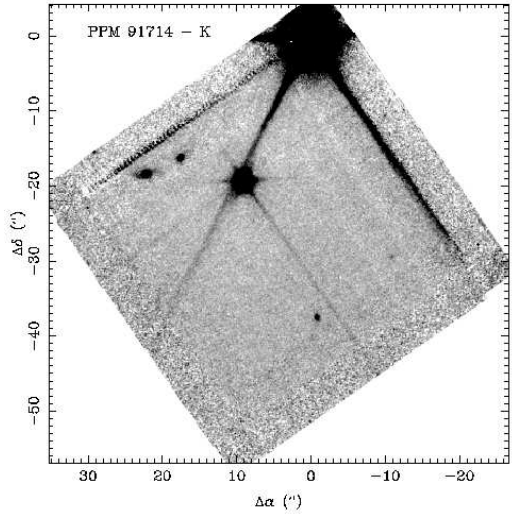}{2.4in}{0}{40}{40}{-120}{-70}
\caption{K band image of the field around ppm 91714.}
\end{figure}

In this paper, we have only identified many sigma objects which are
clearly real in each field. Table 2 lists the galaxies and gives the
galaxy name (specified by the star it is near followed by its relative
RA and Dec in arcseconds), average FWHM, infrared magnitudes and
angular separation of its guide star.  In some of the fields, there
were objects that were difficult to identify as either stars or
galaxies; in these cases we included all the objects in the list and
marked the ones that are ambiguous with a superscript 'a'.

\begin{figure}[htb]
\plotfiddle{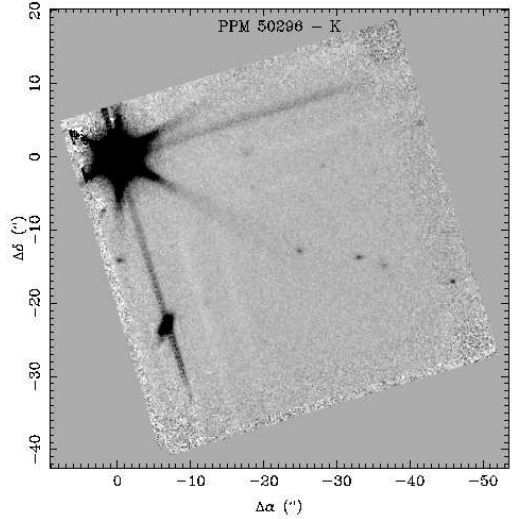}{2.4in}{0}{40}{40}{-120}{-70}
\caption{K band image of the field around ppm 50296.}
\end{figure}

\begin{figure}[hbt]
\plotfiddle{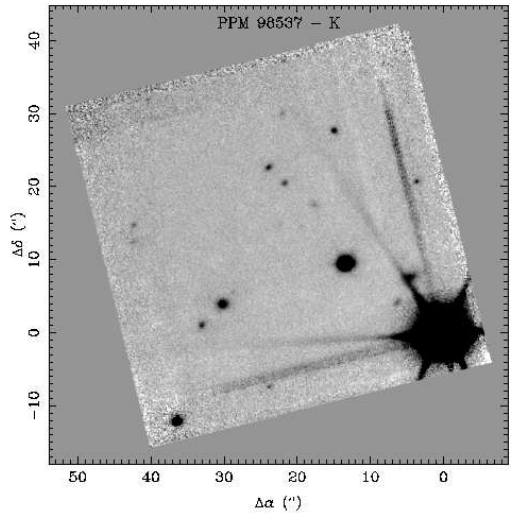}{2.4in}{0}{40}{40}{-120}{-70}
\caption{K band image of the field around ppm 98537.}
\end{figure}

The brightest confirmed galaxy is 16.9 mag in K with a FWHM of
1\farcs53.  The faintest objects identified in each field were
typically $\sim$21st magnitude in K.  Crude morphological types of some
of the galaxies can be determined from these observations, including a
few galaxies with clear spiral structures.  These spiral galaxies are
identified with a superscript 'b' in the table.

\begin{figure}[hbt]
\plotfiddle{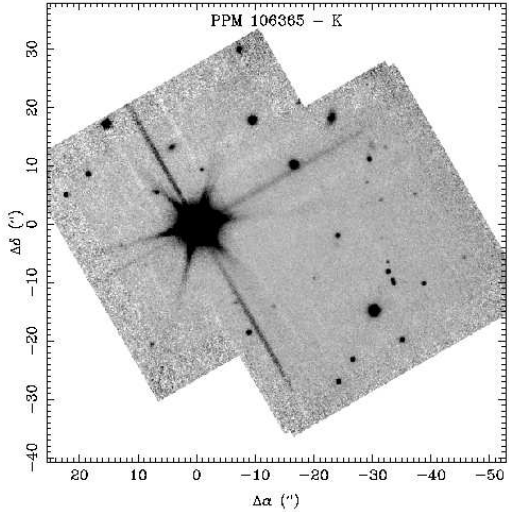}{2.4in}{0}{40}{40}{-120}{-70}
\caption{K band image of the field around ppm 106365.}
\end{figure}

\subsection{PPM 91088}

The field around PPM~91088 is the richest field in terms of resolved
galaxies.  It has at least 20 galaxies between 17th and 21.5 magnitude
(K band) within the slightly less than one square arcminute that is
covered by our images.  These include four bright disk galaxies:
PPM~91088+08+29 (K=19.2 mag and FWHM=0\farcs94), PPM~91088+01+20
(K=18.5 mag and FWHM=0\farcs72), PPM~91088-21+18 (K=17.6 mag and
FWHM=0\farcs83), and PPM~91088-26+09 (K=19.3 mag and FWHM=1\farcs15).
The number density is about 50\% more than would be expected on
average and may indicate weak clustering in this field.

Many of the most interesting galaxies in this field are approximately
20'' to 30'' away from the guide star.  This is not optimal, but it is
alleviated in part by the presence of a relatively bright star
(PPM~91088-23+12, K=16.2 mag) located at a separation of 26\farcs0.  This
star is actually very close to two of the disk galaxies and an AO
camera with a field of view on the order of 10'' should be able to
simultaneously image the psf star and both galaxies.  This would allow
for very accurate psf determinations and deconvolutions.

\subsection{PPM 91714}

This field is relatively empty, except for two bright, potentially
interacting galaxies and one bright star.  The galaxies
(PPM~91714+22-19 and PPM~91714+18-16) are roughly 6'' apart and the
larger of the pair has an asymmetric disk with a full extent of about
5''.  The 2nd galaxy, PPM~91714+18-16, is fainter at K=18.1 mag and
also shows an asymmetrical extension that points away from the
brighter galaxy.  If this does represent an interacting system, one
might expect to see enhanced star formation potentially in the form of
giant star forming regions that could be very compact.

The presence of a bright psf star (PPM~91714+09-19, K=12.6 mag) with
a separation comparable to that of the bright galaxies, makes this an
efficient field to study.  The guide star is also quite bright at R=8.5 mag.
Two other fainter objects are located south of the guide star.

\subsection{PPM 50296}

This is a fairly typical field with seven galaxies brighter than 21.5 mag
at K within the field of view.  But the field is very notable for the
presence of a 16.9 mag disk galaxy; the brightest in any of our five
fields.  The galaxy (PPM 50296-07-23) is highly inclined and has an extent
of about 3''. More significantly, it asymmetrical and has a faint
companion (not necessarily physically associated) about 3'' to the 
south-east.

\subsection{PPM 98537}

PPM~98537 is at a galactic latitude of 28 degrees, and Galactic stars
significantly ``contaminate'' the field. Nevertheless, there is 
at least one identifiable galaxy in the field (PPM~98537+05+08, 
K=19.2 mag) which is quite close to the guide star (offset=8.8'') and
several fainter objects at separations around 20''.

\subsection{PPM 106365}

This is the lowest Galactic latitude field in our sample (23 degrees)
and it is clearly dominated by Galactic stars.  This has one very
positive effect in that there are many stars that can provide accurate
psf's simultaneously with each galaxy image.  There is of course
one negative effect as well, that without very good seeing, it is
difficult to identify which objects are very compact galaxies.

\begin{deluxetable}{lccccc}
\small
\tablewidth{0pt}
\tablecaption{\bf List of Field Galaxies and Stars}
\tablehead{ \colhead{Object Name} & \colhead{FWHM} & \colhead{K Mag} &
\colhead{H Mag} & \colhead{J Mag} & \colhead{Separation('')}}
\startdata

ppm91088+14+13  &      0.53 & 20.0$\pm$0.2 &      $$      &      $$     
 &       18.6 \nl
ppm91088+10+26  &      0.60 & 20.2$\pm$0.2 &      $$      &      $$     
 &       28.3 \nl
ppm91088+08+29 \tablenotemark{b} &      0.94 & 19.2$\pm$0.1 &      $$      &      $$     
 &       30.0 \nl
ppm91088+06+25  &      0.65 & 20.2$\pm$0.1 &      $$      &      $$     
 &       25.3 \nl
ppm91088+01+20 \tablenotemark{b} &      0.72 & 18.5$\pm$0.1 &      $$      &      $$     
 &       20.5 \nl
ppm91088-01+18  &      0.84 & 20.5$\pm$0.2 &      $$      &      $$     
 &       17.8 \nl
ppm91088-03+10 \tablenotemark{a} &      0.58 & 19.3$\pm$0.3 &      $$      &      $$     
 &       10.1 \nl
ppm91088-05+42  &      0.72 & 20.6$\pm$0.2 &      $$      &      $$     
 &       42.5 \nl
ppm91088-08+15  &      1.29 & 20.1$\pm$0.1 &      $$      &      $$     
 &       16.9 \nl
ppm91088-08+45 \tablenotemark{a} &      0.49 & 19.0$\pm$0.1 &      $$      &      $$     
 &       45.7 \nl
ppm91088-13+20 \tablenotemark{a} &      0.61 & 20.6$\pm$0.2 &      $$      &      $$     
 &       23.9 \nl
ppm91088-17+29  &      0.59 & 21.6$\pm$0.3 &      $$      &      $$     
 &       33.9 \nl
ppm91088-19+39  &      0.54 & 20.8$\pm$0.3 &      $$      &      $$     
 &       43.9 \nl
ppm91088-21+18 \tablenotemark{b} &      0.83 & 17.6$\pm$0.1 &      $$      &      $$     
 &       27.7 \nl
ppm91088-22+30  &      0.67 & 18.4$\pm$0.1 &      $$      &      $$     
 &       36.9 \nl
ppm91088-23+12 \tablenotemark{a} & 0.38 & 16.2$\pm$0.1 &      $$      &   $$
 &       26.4 \nl
ppm91088-25-04  &      0.55 & 18.1$\pm$0.1 &      $$      &      $$     
 &       24.9 \nl
ppm91088-25+33  &      0.68 & 21.4$\pm$0.3 &      $$      &      $$     
 &       42.0 \nl
ppm91088-26+09 \tablenotemark{b} &      1.15 & 19.3$\pm$0.1 &      $$      &      $$     
 &       27.4 \nl
ppm91088-26+05  &      0.67 & 19.5$\pm$0.1 &      $$      &      $$     
 &       26.4 \nl
ppm91088-27-11 &      0.57 & 18.5$\pm$0.3 &      $$      &      $$     
 &       29.5 \nl
ppm91088-32+08  &      0.78 & 19.9$\pm$0.1 &      $$      &      $$     
 &       32.8 \nl
ppm91088-35+26  &      0.54 & 19.0$\pm$0.1 &      $$      &      $$     
 &       44.0 \nl
ppm91088-39+14  &      0.50 & 20.3$\pm$0.2 &      $$      &      $$     
 &       41.5 \nl
\hline
ppm91714+22-19 \tablenotemark{b} &      0.62 & 17.0$\pm$0.3 &17.8$\pm$0.3 &18.8$\pm$0.3
 &       28.9 \nl
ppm91714+18-16 \tablenotemark{b} &      0.55 & 18.1$\pm$0.1 &18.3$\pm$0.1 &19.6$\pm$0.1
 &       24.0 \nl
ppm91714+09-19 \tablenotemark{a} &      0.42 & 12.6$\pm$0.3 &12.8$\pm$0.3 &13.7$\pm$0.3
 &       21.4 \nl
ppm91714-01-38  &      0.63 & 19.5$\pm$0.1 &19.9$\pm$0.1 &    $$     
 &       37.8 \nl
ppm91714-11-30  &      0.63 & 20.6$\pm$0.2 &    $$       &    $$     
 &       31.7 \nl
\hline
ppm50296-07-23 \tablenotemark{b} &      1.53 & 16.9$\pm$0.2 &      $$      &      $$     
 &       24.0 \nl
ppm50296-18+01  &      1.13 & 20.3$\pm$0.2 &      $$      &      $$     
 &       17.8 \nl
ppm50296-19-04  &      0.49 & 22.2$\pm$0.6 &      $$      &      $$     
 &       19.0 \nl
ppm50296-25-13  &      0.71 & 20.1$\pm$0.2 &      $$      &      $$     
 &       28.1 \nl
ppm50296-28-01  &      0.63 & 21.3$\pm$0.3 &      $$      &      $$     
 &       28.2 \nl
ppm50296-33-14  &      0.69 & 20.2$\pm$0.2 &      $$      &      $$     
 &       35.9 \nl
ppm50296-37-15  &      0.96 & 20.3$\pm$0.2 &      $$      &      $$     
 &       39.4 \nl
ppm50296-46-17  &      0.51 & 19.9$\pm$0.2 &      $$      &      $$     
 &       49.1 \nl
\hline
ppm98537+43+13  &      0.96 & 20.1$\pm$0.3 &      $$      &      $$     
 &       44.4 \nl
ppm98537+43+15 \tablenotemark{a} &      0.64 & 20.4$\pm$0.3 &      $$      &      $$     
 &       45.1 \nl
ppm98537+37-12 \tablenotemark{a} &      0.64 & 19.2$\pm$0.1 &      $$      &      $$     
 &       38.5 \nl
ppm98537+33+01  &      0.77 & 19.2$\pm$0.1 &      $$      &      $$     
 &       33.2 \nl
ppm98537+30+04  &      0.91 & 17.8$\pm$0.1 &      $$      &      $$     
 &       30.6 \nl
ppm98537+24+23  \tablenotemark{a} & 0.74 & 19.2$\pm$0.1 &      $$      &      $$     
 &       33.2 \nl
ppm98537+22+30  &      0.78 & 20.5$\pm$0.2 &      $$      &      $$     
 &       37.6 \nl
ppm98537+22+21 \tablenotemark{a} & 0.73 & 19.6$\pm$0.1 &      $$      &      $$     
 &       30.2 \nl
ppm98537+18+18  &      1.03 & 20.1$\pm$0.2 &      $$      &      $$     
 &       25.2 \nl
ppm98537+15+28 \tablenotemark{a} & 0.69 & 19.0$\pm$0.1 &      $$      &      $$     
 &       31.7 \nl
ppm98537+06+04  &      0.96 & 20.7$\pm$0.3 &      $$      &      $$     
 &       7.5 \nl
ppm98537+05+08 \tablenotemark{b} &      1.85 & 19.3$\pm$0.3 &      $$      &      $$     
 &       8.8 \nl
\hline
ppm106365+23+05 \tablenotemark{a} &      0.45 & 18.7$\pm$0.1 &      $$      &      $$     
 &       23.2 \nl
ppm106365+19+09 \tablenotemark{a} &      0.47 & 18.3$\pm$0.1 &      $$      &      $$     
 &       20.7 \nl
ppm106365+16+17 \tablenotemark{a} &      0.47 & 15.8$\pm$0.2 &      $$      &      $$     
 &       23.4 \nl
ppm106365+08-21 &      0.58 & 20.7$\pm$0.4 &      $$      &      $$     
 &       22.1 \nl
ppm106365+07+06 &      0.61 & 19.6$\pm$0.1 &      $$      &      $$     
 &       9.0 \nl
ppm106365+04+13 &      0.76 & 18.4$\pm$0.1 &      $$      &      $$     
 &       14.1 \nl
ppm106365-01+10 \tablenotemark{a} &      0.43 & 20.0$\pm$0.2 &      $$      &      $$     
 &       9.5 \nl
ppm106365-07-13 &      0.79 & 20.4$\pm$0.1 &      $$      &      $$     
 &       15.0 \nl
ppm106365-07+30 \tablenotemark{a} &      0.50 & 17.7$\pm$0.2 &      $$      &      $$     
 &       31.1 \nl
ppm106365-09-19 \tablenotemark{a} &      0.45 & 18.2$\pm$0.3 &18.4$\pm$0.3 &19.0$\pm$0.3
 &       20.5 \nl
ppm106365-09+18 \tablenotemark{a} &      0.48 & 16.0$\pm$0.1 &      $$      &      $$     
 &       20.4 \nl
ppm106365-13-14 \tablenotemark{a} &      0.59 & 20.5$\pm$0.2 &21.6$\pm$0.4 &23.2$\pm$0.6
 &       19.2 \nl
ppm106365-17+10 \tablenotemark{a} &      0.48 & 16.0$\pm$0.3 &16.1$\pm$0.3 &16.8$\pm$0.3
 &       19.6 \nl
ppm106365-20-09 \tablenotemark{a} &      0.42 & 20.7$\pm$0.2 &20.8$\pm$0.2 &21.5$\pm$0.2
 &       22.0 \nl
ppm106365-23+18 \tablenotemark{b} &      1.21 & 17.7$\pm$0.1 &18.6$\pm$0.1 &19.6$\pm$0.1
 &       29.5 \nl
ppm106365-24-27 \tablenotemark{a} &      0.44 & 18.2$\pm$0.1 &18.2$\pm$0.1 &18.8$\pm$0.1
 &       36.3 \nl
ppm106365-24-02 \tablenotemark{a} &      0.46 & 18.6$\pm$0.1 &18.8$\pm$0.1 &19.3$\pm$0.3
 &       24.1 \nl
ppm106365-27-23 \tablenotemark{a} &      0.46 & 18.4$\pm$0.1 &18.5$\pm$0.1 &19.3$\pm$0.1
 &       35.4 \nl
ppm106365-29+07 \tablenotemark{a} &      0.50 & 21.1$\pm$0.3 &21.2$\pm$0.2 &22.1$\pm$0.2
 &       30.0 \nl
ppm106365-29+11 &      0.60 & 18.8$\pm$0.3 &19.5$\pm$0.3 &20.6$\pm$0.3
 &       31.6 \nl
ppm106365-30-15 \tablenotemark{a} &      0.45 & 14.9$\pm$0.1 &15.0$\pm$0.1 &15.5$\pm$0.1
 &       33.7 \nl
ppm106365-30+14 &      1.07 & 19.7$\pm$0.3 &20.3$\pm$0.3 &21.1$\pm$0.3
 &       32.9 \nl
ppm106365-31+04 \tablenotemark{a} &      0.51 & 20.4$\pm$0.2 &21.3$\pm$0.2 &21.8$\pm$0.2
 &       31.8 \nl
ppm106365-32+13\tablenotemark{a}  &      0.55 & 21.4$\pm$0.7 &21.3$\pm$0.4 &23.2$\pm$0.5
 &       34.9 \nl
ppm106365-33-06 \tablenotemark{a} &      0.44 & 21.3$\pm$0.4 &20.6$\pm$0.3 &21.2$\pm$0.3
 &       33.3 \nl
ppm106365-33-08 \tablenotemark{a} &      0.47 & 18.6$\pm$0.3 &18.6$\pm$0.3 &18.9$\pm$0.3
 &       33.7 \nl
ppm106365-35-20 \tablenotemark{a} &      0.45 & 18.3$\pm$0.1 &18.4$\pm$0.1 &19.0$\pm$0.1
 &       40.3 \nl
ppm106365-37+05 \tablenotemark{a} &      0.48 & 20.9$\pm$0.3 &21.4$\pm$0.3 &22.1$\pm$0.3
 &       37.7 \nl
ppm106365-39-10 \tablenotemark{a} &      0.48 & 19.3$\pm$0.1 &19.4$\pm$0.1 &20.0$\pm$0.1
 &       40.2 \nl
ppm106365-46-06 \tablenotemark{a} &      0.48 & 20.4$\pm$0.2 &20.3$\pm$0.2 &21.4$\pm$0.2
 &       46.2 \nl

\enddata
\tablenotetext{a}{unresolved, possible star}
\tablenotetext{b}{apparent spiral structure}
\end{deluxetable}

\clearpage
\begin{figure}
\plotfiddle{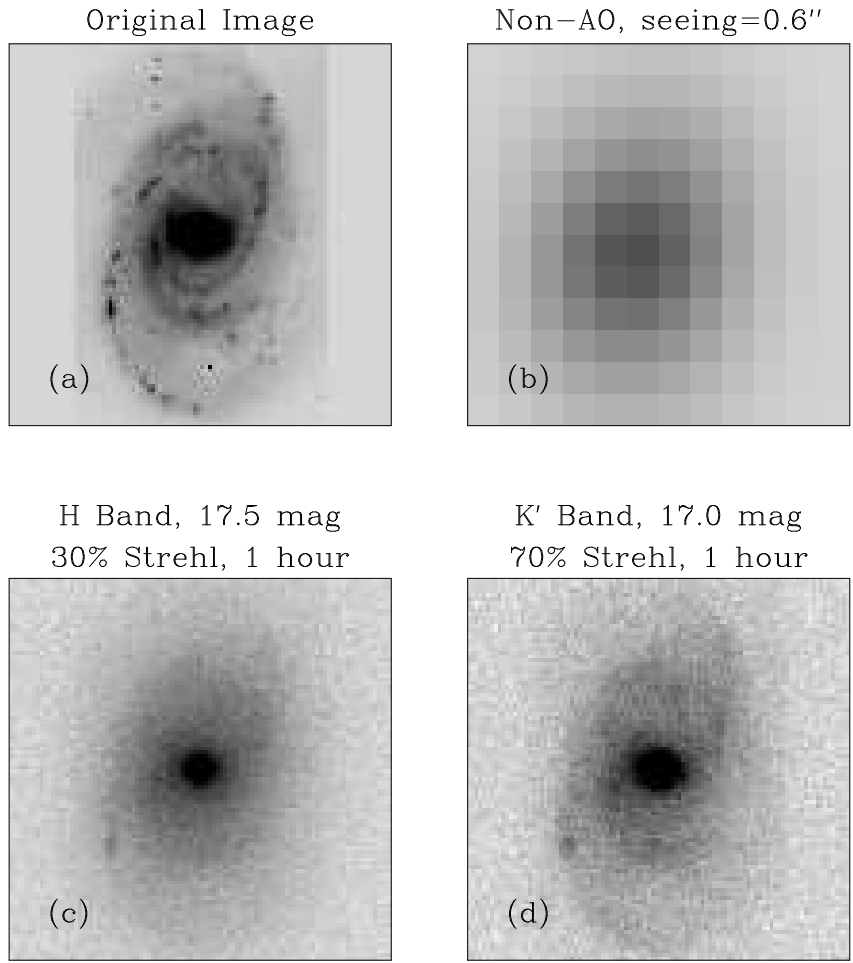}{4in}{0}{80}{80}{-230}{-130}
\caption{Simulated images of a large spiral galaxy. Panel (a) shows the
original image which is a reduced version of an R band image of NGC~5371.
The resolution is essentially 1 pixel due to resampling a much larger
original image.  Panel (b) shows the result of blurring the original
image with a 0.6'' seeing disk.  Panel(c) shows a simulated
undeconvolved image of the galaxy taken in the H band with 30\% Strehl.
The image was scaled in intensity to match a 17.5 mag galaxy and
represents a 1 hour integration.  Panel (d) is a similar simulation
for the K band.  Notice that in the AO simulated frames, the disk is
easily distinguished from the central bulge and spiral arms
are detectable.  Also the brightest star forming knot to the lower
left of the galaxy is also apparent.
}
\end{figure}

\clearpage
\section{Galaxy Simulations}

An important concern in imaging faint galaxies is sensitivity.  In
particular, many of these galaxies are difficult to image when most of
their light is concentrated in a few pixels, how much more difficult
will this be when they are sampled at 0\farcs02 per pixel?  Also for
morphological studies, you want not only to detect the galaxy but also
measure its brightness over some extended area, or at least determine
its size and light profile.  Thankfully, the background per pixel also
goes down as the square of the pixel scale, so these studies are
possible even in the near infrared.

\begin{figure}[hbt]
\plotfiddle{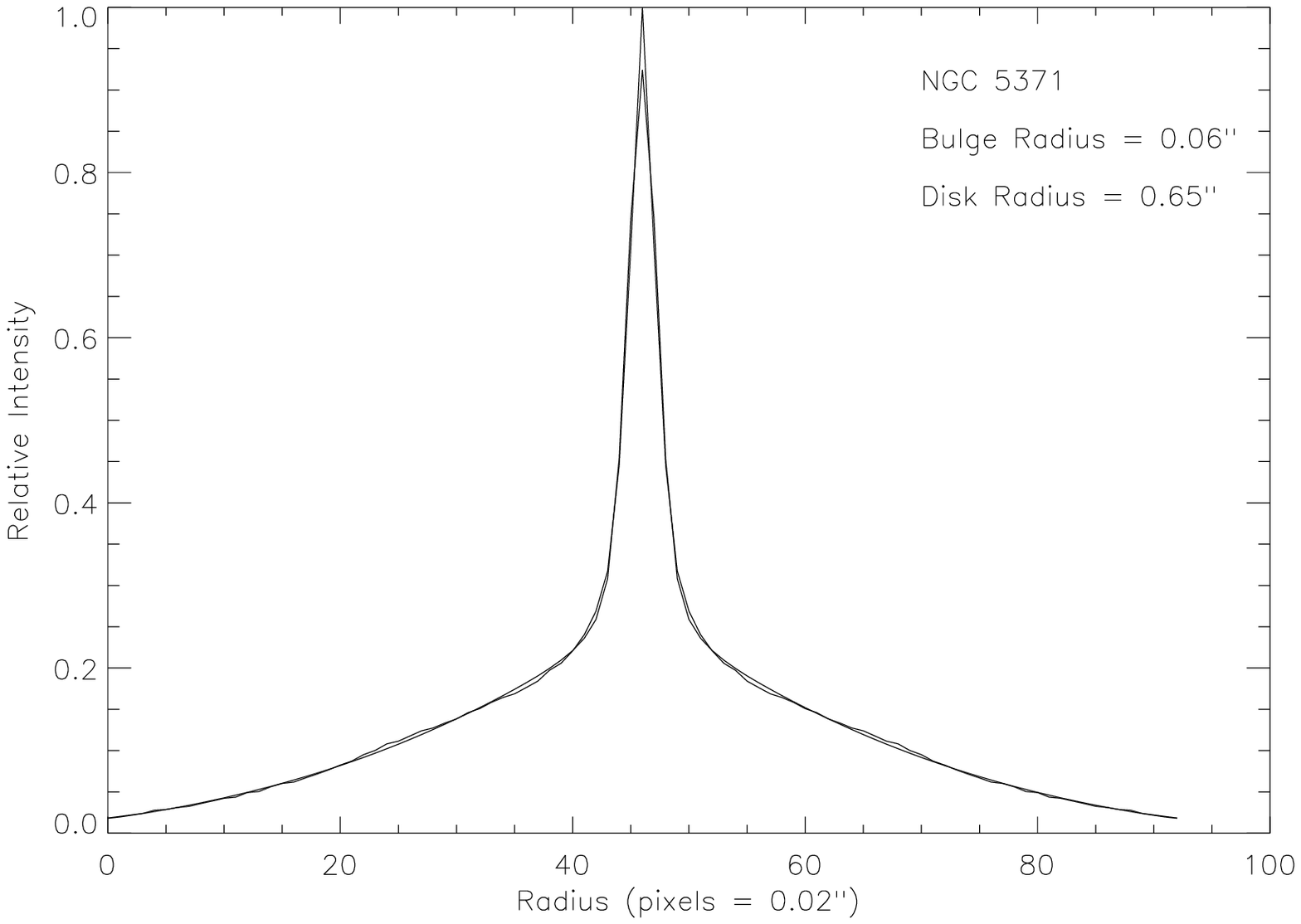}{1.6in}{0}{40}{40}{-130}{-150}
\caption{The azimuthally averaged radial profile of the simulated AO
image based on NGC~5371. The smoother curve is a theoretical fit with
a de Vaucouleur's profile for the bulge and an exponential disk.  The
effective radii are well determined with r$_{bulge}$=0\farcs06 and
r$_{disk}$=0\farcs65.  }
\end{figure}

\begin{figure}
\plotfiddle{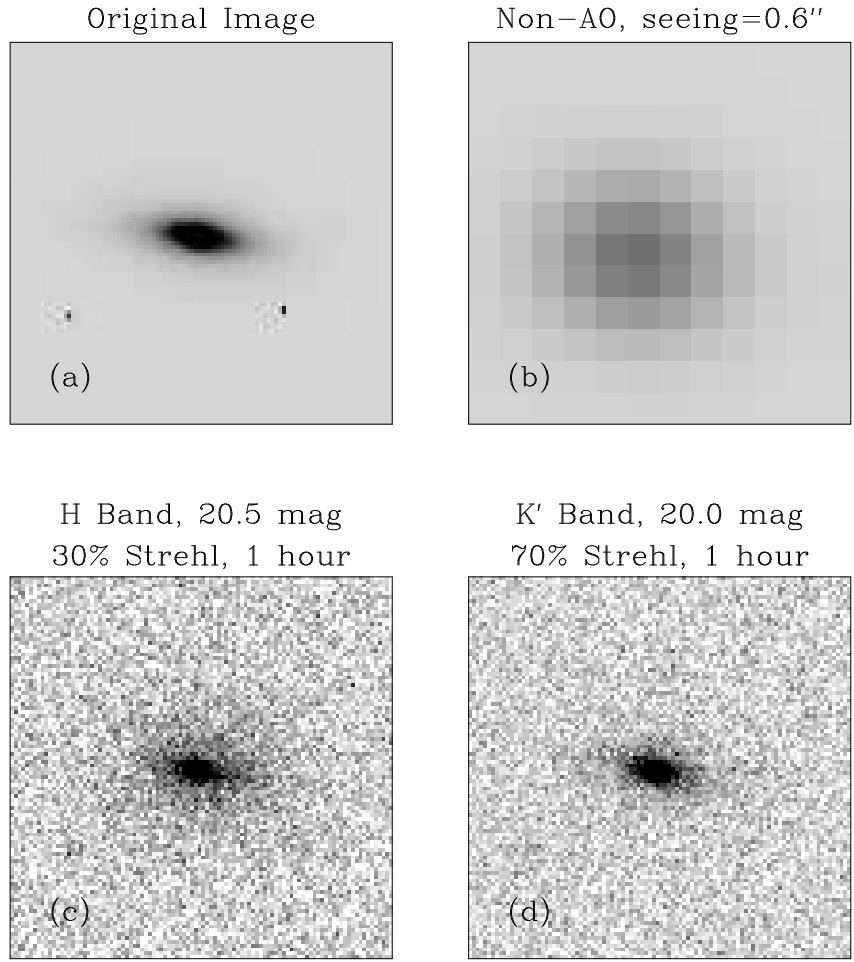}{4in}{0}{80}{80}{-230}{-130}
\caption{Simulated images of a faint compact galaxy similar to figure 6.
Panel (a) shows the original image of NGC~4036.
The resolution is essentially 1 pixel due to resampling a much larger
original image.  Panel (b) shows the unresolved image that
is present in any good seeing (0\farcs6) non-AO image.  Panels (c)
and (d) show the AO simulations at H and K' band respectively. In
both panels, the tiny galaxy is easily resolved in both axes, so an
accurate size can be determined.
}
\end{figure}

\begin{figure}[hbt]
\plotfiddle{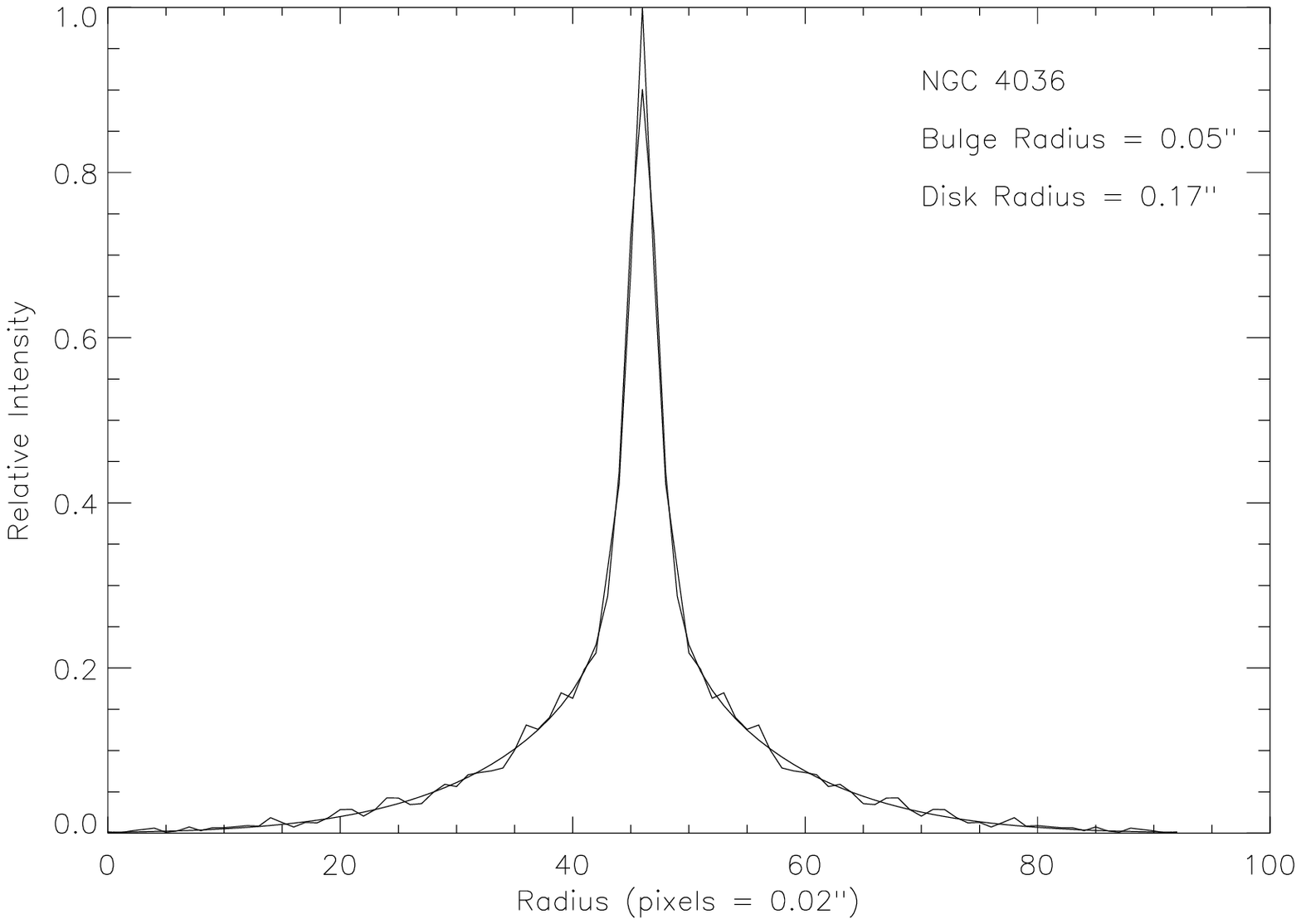}{1.6in}{0}{40}{40}{-130}{-150}
\caption{The azimuthally averaged radial profile of the simulated AO
image based on NGC~4036. The smoother curve is a theoretical fit with
a de Vaucouleur's profile for the bulge and an exponential disk. The
effective radii are well determined with r$_{bulge}$=0\farcs05 and
r$_{disk}$=0\farcs15. }
\end{figure}

To quantify this, high quality R band images of the nearby spiral
galaxy NGC~5371 and the S0 galaxy NGC~4036 were used to create
artificial AO images.  The R band was selected because it is roughly
red shifted to the H band at a redshift of about 1.5, where we might
expect to find a significant number of faint galaxies.  All
simulations assume the Keck Telescope (10~m) with a one hour
integration and a camera throughput of 30\%; significantly worse than
the current non-AO near infrared camera (NIRC, $\sim$46\% from
Matthews \& Soifer \markcite{mat} 1994).  It is also assumed that the
object can be dithered on chip to generate sky measurements. The
simulated backgrounds were 13.7 at H and 12.9 at K' corresponding to
the nominal H band sky background at Mauna Kea, but an increased K'
background (nominal is 13.9) in order to simulate additional thermal
emission from the AO system.  The Strehl ratio was assumed to be 30\%
at H with a diffraction limited fwhm of 0\farcs04 and a halo with a
fwhm of 0\farcs6. For K band, the Strehl was 70\% with a core fwhm of
0\farcs06 and a halo of 0\farcs6.  No deconvolutions were performed on
any of the images.  Since some galaxies are available with small
separations from their potential guide star, no anisoplanatic effects
are included.  This clearly becomes an important factor as the Strehl
ratio declines rapidly beyond 20-30 arcseconds.

NGC~5371 was selected to compare to the resolved galaxies that make up
the brightest members of our sample (e.g. PPM~91088-21+18,
PPM~91088-25-04, PPM~91714+22-19, PPM~50296-07-23, PPM~106365-23+18).
It has a bright central halo and a near face-on spiral disk.  The
galaxy was resampled onto a 100x100 grid simulating a 2'' field of
view with 0\farcs02 /pixel.  The visible disk was given an extent of
roughly 2''x1'' comparable again to our brightest candidates.  The
total magnitude was set to 17.0 at K' and 17.5 in H band.  Figure 6
shows 4 panels of NGC~5371 under different conditions. Panel (a) is
the original image with no noise and essentially 1 pixel resolution.
Panel (b) is a simulated image under good seeing conditions (0\farcs6)
but no AO system and no noise. It has a plate scale of 0\farcs16 per
pixel.  This panel shows essentially an unresolved object since the
central region dominates the light distribution.  Panel (c) is the H
band simulated image with the AO parameters described in the last
paragraph.  One bright HII region is visible to the bottom left of
the disk, and the disk is easily observed.  Little of the spiral
structure is apparent in this raw image.  Panel (d) is the K
band simulation. Both the bright HII region and the spiral arms
are easily distinguished.  The core is elongated, but not quite
enough to distinguish the small central bar present in this galaxy.
Figure 7 shows an azimuthally averaged radial profile of the galaxy.
The smoother curve is a theoretical fit with a de Vaucouleur's profile
for the bulge and an exponential disk.  The effective radii are well
determined with r$_{bulge}$=0\farcs06 and r$_{disk}$=0\farcs65. Notice
that the noise in the radial plot is extremely low even though the
signal to noise in each pixel is quite modest.

The second galaxy (NGC~4036) was selected to test how well the size
and basic morphology could be determined for some of our faintest and
smallest candidates.  The image was resampled onto a 100x100 pixel
grid (0\farcs02 per pixel) such that its horizontal fwhm was 0\farcs1 and
its vertical fwhm was 0\farcs08.  Its flux was scaled to a 20th magnitude
object at K and 20.5 at H.  Figure 8 shows the simulated galaxy under
four different conditions.  Panel (a) is the original with 1 pixel
resolution and no noise.  Panel (b) shows the unresolved image that
is present in any good seeing (0\farcs6) non-AO image.  Panels (c)
and (d) show the AO simulations at H and K' band respectively.
Figure 9 shows an azimuthally averaged radial profile of the galaxy.
The smoother curve is a theoretical fit with a de Vaucouleur's profile
for the bulge and an exponential disk.  The effective radii are well
determined with r$_{bulge}$=0\farcs05 and r$_{disk}$=0\farcs15.

\section{Conclusions}

In this paper, we have presented a new strategy for observing faint
compact galaxies with a high order AO system.  Over 40 galaxies were
identified near 5 bright stars, all appropriate candidates for early
Adaptive Optics observations with large ground based telescopes. Our
simulations demonstrate that typical objects found in the fields are
observable and that fundamental galaxy properties such as disk and
bulge size can be measured.  We believe these observations will
greatly facilitate the future diffraction limited observations of
faint field galaxies, even with the very limited fields of view of
early AO cameras.

\acknowledgments The authors are very grateful for the support and
encouragement of Ian McLean and Andrea Ghez.  We would also like to
thank Alycia Weinberger and Bruce Macintosh for many useful conversations
and assistance with observing.  This work would not be possible
without the help and interaction with the adaptive optics team at
Keck: Peter Wizinowich, Scott Acton, Olivier Lai, Chris Shelton, and
Paul Stomski.  Finally we would like to thank our telescope operator,
Gary Puniwae, and all of the Keck staff.


\end{document}